\documentclass[prx,twocolumn,groupedaddress]{revtex4-1}
\usepackage{graphicx}
\usepackage{dcolumn}
\usepackage{bm}

\usepackage{amsmath}
\usepackage{graphicx}
\usepackage{color}
\usepackage{wrapfig}
\usepackage{epstopdf}

\graphicspath{{pics/}}

\newcommand{\om}{\omega}
\newcommand{\bee}{\begin{equation}}
\newcommand{\ene}{\end{equation}}

\newcommand{\bea}{\begin{eqnarray}}
\newcommand{\ena}{\end{eqnarray}}

\begin{document}

\title{Large Purcell enhancement without strong field localization}

\author{Alexander Krasnok,$^{1}$ Stanislav Glybovski,$^{1}$ Mihail Petrov,$^{1}$ Sergey Makarov,$^{1}$\\ Roman Savelev,$^{1}$ Pavel Belov$^{1}$, Constantin Simovski$^{1,2}$, and Yuri Kivshar$^{1,3}$}
\address{
$^{1}$ITMO University, St.~Petersburg 197101, Russia\\
$^{2}$Aalto University, School of Electrical Engineering
P.O. Box 13000, 00076 Aalto, Finland\\
$^{3}$Nonlinear Physics Center, Australian National University, Canberra ACT 0200, Australia}

\begin{abstract}
The Purcell effect is defined as the modification of spontaneous decay in the presence of a resonator, and in plasmonics it is usually associated with the large local-field enhancement in "hot spots" due to surface plasmon polaritons. Here we propose a novel strategy for enhancing the Purcell effect through engineering the radiation directivity without a strict requirement of the local field enhancement. Employing this approach, we demonstrate how to enhance the Purcell effect by two orders of magnitude in all-dielectric nanostructures recently suggested as building blocks of low-loss nanophotonics and metamaterials. We support our concept by proof-of-principle microwave experiments with arrays of high-index dielectric resonators.
\end{abstract}

\maketitle

\section{Introduction}

The Purcell effect is known as a modification of the spontaneous emission lifetime of a quantum emitter induced by its interaction with environment~\cite{Purcell_46,Sauvan2013a}. This modification is significant if the environment is a resonator tuned to the emission frequency. Although the Purcell effect was discovered in the context of nuclear magnetic resonance~\cite{Purcell_46}, nowadays it is widely used in many applications, ranging from microcavity light-emitting devices~\cite{Fainman_2010} to single-molecule optical microscopy~\cite{Koenderink_PRL_11, Cosa_2013}, being also employed for tailoring optical nonlinearities~\cite{Soljacic_2007} and enhancement of spontaneous emission from quantum cascades~\cite{Minot_2007}. Open nanoscale resonators such as plasmonic nanoantennas can change the spontaneous emission lifetime of a single quantum emitter, that is very useful in microscopy of single NV centers in nanodiamonds~\cite{Vamivakas_NanoLetters_13}, Eu$^{3+}$-doped nanocrystals~\cite{Carminati_PRL_14}, plasmon-enhanced optical sensing~\cite{Sauvan2013a}, and the visualization of biological processes with large molecules~\cite{Tinnefeld_Science_2012}.

It is generally believed that high values of the Purcell factor can be achieved in the systems with the strong local-field enhancement associated with the formation of "hot spots" (e.g. in nanoantennas)~\cite{Koenderink2010, Sauvan2013a}. Accordingly, in order to achieve high Purcell factor ($F\gg1$) a quantum emitter should be placed in one of such hot spots. This strategy follows from the well-known formula for $F$~\cite{Novotny_Hecht_book},
\begin{equation}\label{eq0}
{\rm F}=1+\frac{6\pi\varepsilon_0\varepsilon}{k^{3}|{\bf d}|^{2}}\mbox{Im}\left[ {\bf d}^{*}\cdot{\bf E}_s({\bf R}_0)\right],
\end{equation}
where $k$ is the wave number, $\mathbf{d}$ is the radiating electric dipole moment of an emitter, and $\mathbf{E}_s(\mathbf{R}_0)$ is a scattered electric field at the emitter origin $\mathbf{R}_0$ produced by a resonator or nanoantenna, $\varepsilon_0$ and $\varepsilon$ are the dielectric constants of vacuum and surrounding media. The field produced by an emitter is much smaller than the internal field ${\bf E}_0$, with the local field ${\bf E}_s$ compensating the reactive part of ${\bf E}_0$. This implies that a plasmonic nanoantenna generates a hot spot covering the emitter. An increase of the local-field intensity due to the presence of a nanoantenna can be described by the so-called Local-Field Enhancement Factor (LFEF) $\Sigma\equiv |{\bf E}_s+{\bf E}_0|^2/|{\bf E}_0|^2$ which attains the values of $\Sigma \sim 10^2\dots 10^3$ \cite{Barrow2014, Koenderink2010}. Usually, LFEF is calculated when the plasmonic nanoantenna is excited not by the near field of the emitter but by an external wave at the resonant frequency. Then the same resonant mode is excited in the structure as in the case of  spontaneous emission. Though LFEF calculated for this scattering case does not coincide with to LFEF calculated for the case of spontaneous emission, these values are of the same order~\cite{Sauvan2013a}.

When we replace metallic nanoantennas with dielectric structures~\cite{Cummer_08,Brener_12,Lukyanchuk13}, we expect somewhat qualitatively similar physics to apply. Moreover, the recent progress in resonant all-dielectric nanophotonics has opened a door to many important applications of such structures including sensors and nanoantennas~\cite{Miroshnichenko:NL:2012}, nanowaveguides~\cite{Savelev2014_1}, Huygens' metasurfaces~\cite{Staude_15}, and metamaterials~\cite{Cummer_08, Brener_12}. However, all-dielectric nanostructures can not provide high values of the local fields even in hot spots~\cite{Kuznetsov15}, and they are believed to demonstrate much low values of the Purcell factor~\cite{Asano2007,Schmidt_12, KrasnokOE, Schuller:OE:2009}, thus not supporting efficient interaction of nanostructures with light emitters.

In this paper, we suggest and demonstrate a novel methodology for tailoring the Purcell factor in terms of a local-field enhancement and radiation directivity of radiation, and arrive at somewhat counterintuitive conclusion that large Purcell enhancement can be achieved even in radiating all-dielectric systems without strong local field localization. Using this approach, we predict and demonstrate in the proof-of-principle microwave experiments large values of the Purcell factor in all-dielectric nanostructures near the frequency of the optically-induced magnetic resonant response.

\begin{figure}[!t]
\includegraphics[width=0.48\textwidth]{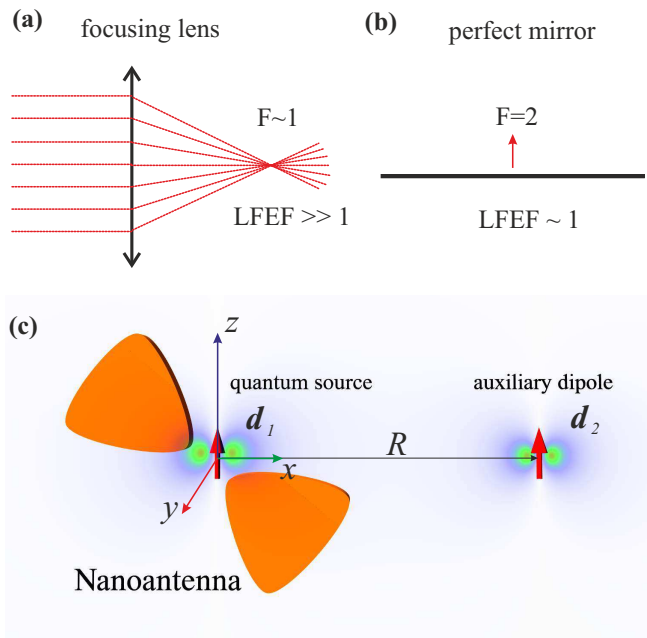}
\caption{(a) System with a high local-field enhancement factor (LFEF), but without Purcell effect (focusing lens). (b) System without LFEF, but has the Purcell effect (thin perfect mirror). (c) A schematic illustration of a reciprocity theorem applied to a single emitter with dipole moment ${\bf d}_1$ (emitter). Dipole ${\bf d}_2$ is an auxiliary. The emitter ${\bf d}_1$ is located near an arbitrary nanostructure (e.g. nanoantenna). The distance between ${\bf d}_1$ and ${\bf d}_2$ is equal $R$.}\label{esciz}
\end{figure}

We mention that the opposite examples are more obvious, and large LFEF is not always associated with large Purcell factor of an emitter located at the same point. Intuitively clear example is a huge LFEF in a focal point of optical focusing lens [see Fig.~\ref{esciz}(a)]. The Purcell factor of the emitter located at a focal point is unity due to a large distance from the lens. An example of a weak LFEF but with a large Purcell factor is shown in Fig.~\ref{esciz}(b), where the Purcell factor of a dipole orthogonal to a perfect mirror surface is 2~\cite{Sipe_1984}. A thin perfect mirror surface has LFEF$=1$ under irradiation by a p-polarized plane wave, but it changes greatly the radiation pattern of an emitter. This example suggests some relation between the Purcell factor and the radiation directivity, that we study below in more detail.

\section{Engineering the Purcell factor}

First, we follow the earlier approach~\cite{Novotny_09_AOP} and discuss how to engineer the Purcell effect through the radiation directivity. We consider the geometry shown schematically in Fig.~\ref{esciz}(c), with a plasmonic or dielectric nanoantenna,
an emitter with dipole moment ${\bf d}_1$ placed near the nanostructure, and also an auxiliary dipole ${\bf d}_2$ placed at the distance $R$ from the nanoantenna. The auxiliary dipole is required to use the reciprocity theorem, and the final results do not contain the dipole moment ${\bf d}_2$. Initially, we assume that the dipole ${\bf d}_2$ is a emitter and ${\bf d}_1$ is a receiver of light. The power absorbed by the dipole ${\bf d}_1$ from a radiation field of the dipole ${\bf d}_2$ (taking into account a re-scattered field from the nanostructure), can be written as~\cite{Novotny_09_AOP}:
\begin{equation}\label{eq1}
P_{\rm ex}=\frac{\omega}{2}\mbox{Im}\left[\mathbf{d}_1^*\cdot \mathbf{E}_2(0)\right],
\end{equation}
where $\mathbf{E}_2(0)$ is the total electric field at a point of the dipole ${\bf d}_1$ which includes the field of the dipole ${\bf d}_2$ and the re-scattered field from the nanostructure. We denote the unit vectors of the first and second dipoles as $\mathbf{e}_1$ and $\mathbf{e}_2$, respectively. Then, the dipole moment of the first (second) dipole can be re-written as $\mathbf{d}_{1,2}=\alpha_{1,2} (\mathbf{e}_{1,2}\otimes \mathbf{e}_{1,2})\cdot\mathbf{E}_{2,1}$, where $\alpha_{1,2}$ is a polarizability of the first (second) dipole, $\otimes$ is a tensor product. Therefore, the expression (\ref{eq1}) can be rewritten as $P_{\rm ex}=\omega/2\cdot\mbox{Im}\left[\alpha_1\right]\left|\mathbf{e}_{1}\cdot \mathbf{E}_{2}(0)\right|^2$. In this case the reciprocity theorem can be represented as~\cite{Novotny_09_AOP}:
\begin{equation}\label{eq2}
\mathbf{d}_1\mathbf{E}_{2}(0)=\mathbf{d}_2\mathbf{E}_{1}(\mathbf{R}).
\end{equation}
Then using Eq.~(\ref{eq2}) one can rewrite Eq.~(\ref{eq1}) in the following form: $P_{\rm ex}=\omega/2\cdot\mbox{Im}\left[\alpha_1\right]\left|d_2/d_1\right|^2\cdot\left|\mathbf{e}_{2}\cdot \mathbf{E}_{1}(\mathbf{R})\right|^2$, where $\mathbf{E}_{1}(\mathbf{R})$ is the field created by the dipole $\mathbf{d}_1$ in the presence of the nanostructure at the point of the dipole $\mathbf{d}_2$. In this expression, the quantity $\left|\mathbf{e}_{2}\cdot \mathbf{E}_{1}(\mathbf{R})\right|^2$ is proportional to the power flow associated with the field polarization along $\mathbf{e}_{2}$, which is radiated by a dipole emitter $\mathbf{d}_{1}$ at the location of the dipole $\mathbf{d}_{2}$. Then, the dipole $\mathbf{d}_{2}$ can be moved to the farfield zone of the radiation ($R\gg\lambda$, $\lambda=2\pi c/\omega$ -- emission wavelength). In this case the field $\mathbf{E}_{1}(\mathbf{R})$ is fully transverse. We can introduce the radiation directivity with polarisation along $\mathbf{e}_{2}$, as $D_e=4\pi\left|\mathbf{e}_{2}\cdot \mathbf{E}_{1}(R,\theta,\varphi)\right|^2/P_{\rm rad}$, where $P_{\rm rad}=\int{\left|\mathbf{E}_{1}(R,\theta,\varphi)\right|^2}d\Omega$ is a total radiated power of the emitter. Introducing this quantity, we arrive at the expression $P_{\rm ex}=\omega/(8\pi)\mbox{Im}\left[\alpha_1\right]\left|d_2/d_1\right|^2 D_e(R,\theta,\varphi) P_{\rm rad}$, that defines the power absorbing by dipole $\mathbf{d}_1$ located in the vicinity of a nanostructure, as a function of its radiated power $P_{\rm rad}$ and directivity $D_e(R,\theta,\varphi)$. In the approximation of the dipole moment independence on the presence of nanostructure, the same expression for the dipole $\mathbf{d}_1$ located in a free space can be written as: $P^{0}_{\rm ex}=\omega/(8\pi)\mbox{Im}\left[\alpha_1\right]\left|d_2/d_1\right|^2 D^{0}_e(R,\theta,\varphi) P^{0}_{\rm rad}$. The all values with index ''0" refer to the free space. The ratio of the obtained expressions with and without nanostructure leads to the following formula:
\begin{equation}\label{eq3}
\frac{P_{\rm ex}(\theta,\varphi)}{P^{0}_{\rm ex}(\theta,\varphi)}=\frac{D_e(\theta,\varphi)}{D^{0}_e(\theta,\varphi)}\cdot\frac{P_{\rm rad}}{P^{0}_{\rm rad}}.
\end{equation}

For our further analysis, the Eq.~(\ref{eq3}) needs to be reformulated as follows. We will study the radiation pattern of the system only in the plane $z=0$ ($\theta=\pi/2$), and will consider only the cases of parallel orientation of the vectors $\mathbf{E}_{1}$ and $\mathbf{e}_{2}$. Thus we consider the ability of rotation of the polarization plane of the emitted light due to the presence of the nanostructure. We denote the local-field enhancement factor (LFEF) as $\Sigma_e$. The quantity $P_{\rm rad}/P^{0}_{\rm rad}$ is equal to the radiation part of the PF, which we denote as $F_{\rm rad}$. Radiation part of the PF related to the total Purcell factor (\ref{eq0}) through radiation efficiency ($\eta$): $F_{\rm rad}=\eta F$. It is well known, the directivity of a dipole in a free space is $D^0(\theta,\varphi)=3/2\sin^2\theta$. But in our case $\theta=\pi/2$, and therefore $\sin^2\theta=1$. Then, we come to the final expression, which connects the radiation part of a PF with a LFEF ($\Sigma_e$) and radiation directivity ($D$):
\begin{equation}\label{eq4}
F_{\rm rad}=\frac{3}{2}\frac{\Sigma^2_e(\varphi)}{D(\varphi)}.
\end{equation}
Due to the reciprocity theorem, the quantities $\Sigma^2(\varphi)$ and $D(\varphi)$ are equally dependent on the angle $\varphi$, and therefore the value $F_{\rm rad}$ is independent on the directions in space. We emphasize that the Eq.~(\ref{eq4}) relates the characteristics of the nanoantenna in receiving and transmitting modes (if our arbitrary structure is a nanoantenna).

Thus, the value of the radiation PF can be expressed in terms of the radiation directivity of the system "emitter $+$ nanostructure" in some direction $\varphi$ and the LFEF at the location of the emitter, under irradiation by a plane wave from the same direction ($\varphi$). The Eq.~(\ref{eq4}) allows to make some general statements about the PF based on some basic principles of the system symmetry. For example, there are approaches to estimate the value of $D(\varphi)$ even for complex radiating systems~\cite{Balanis2005}. From here it follows that a radiating system has the smallest directivity for a symmetric emitter location. Therefore, for the same $\Sigma$ the radiation PF is maximal at the emitter locations close to the symmetry axis of a radiating system.

In addition to the examples shown in Figs.~\ref{esciz}(a) and (b) let us consider a few more. In order to reduce the radiation in same direction, a metal or a dielectric plate can be placed closely to the emitter, or for saving a symmetry -- two mirrors. Thus we come to the quantum emitter in a Fabry-Perot cavity. Talking about the dielectric nanostructures, all--dielectric Yagi-Uda nanoantenna~\cite{KrasnokOE} has low both LFEF ($\Sigma_e$) and high radiation directivity (more than 10). Therefore, such nanoantennas have a low value of PF. In previous works it was usually associated with low values of the LFEF $\Sigma_e$.

All of these examples relate to the cases when the radiation pattern change is accompanied by the increase of  $\Sigma_e$. However, radiating systems which have both a small value of $\Sigma_e$, under irradiation at angle $\varphi$ and a small directivity $D(\varphi)$ of radiation in the same direction are fundamentally possible. Such systems can have a high value of PF, although the LFEF in this point is not strong. One example of such a structure is shown in Fig.~\ref{esciz}(b). This is explained by the presence of so-called \textit{dark eigenmodes}. Dark eigenmodes of nanostructure are the eigenmodes which can not be excited by a plane wave~\cite{magdark_5, Yang2015}. When nanostructure is excited by a dipole emitter, these modes are excited very well and contribute to the radiation directivity. As an example, in a recent study~\cite{KrasnokNanoscale}, the superdirectivity regime of all--dielectric nanoantennas have been achieved in such a way. Thus, even if the LFEF of nanostructure is not very strong, the high PF can be achieved through the designing of a ''special'' directivity pattern. For example, when a quadrupole multipole moment a(l,m) (electric or magnetic type) of order (2,0) is excited in the structure, the radiation in a plane of $\theta=\pi/2$ is suppressed. In this case, such a structure can have infinitely high PF if the LFEF is not equal to zero under irradiation this structure from the same direction. However, actually the set of multipole moments is always excited and value (\ref{eq4}) of PF remains finite.

\begin{figure}[!b]
 \includegraphics[width=0.45\textwidth]{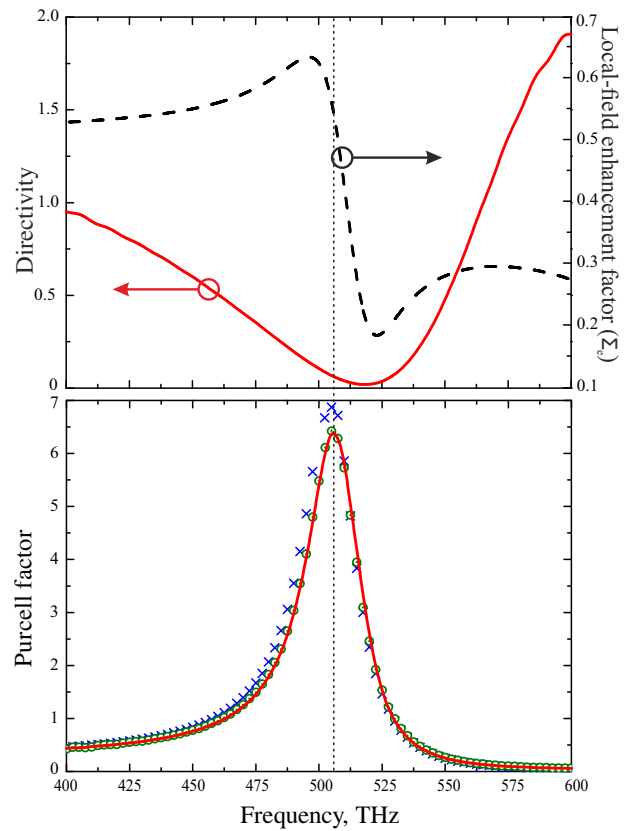}
 \caption{(a) Radiation directivity in ($\theta=\pi/2$, $\varphi=\pi/2$) direction (red solid curve) and local-field enhancement factor (black dashed curve) under irradiation from the same direction, for all-dielectric dimer nanoantenna. (b) The Purcell factor of all-dielectric dimer nanoantenna calculated by the formula (\ref{eq4}) (red curve), by discrete dipole approximation (DDA, blue crosses), and through method of input impedances (green circles)~\cite{Krasnok_Purcell_2015} as dependencies on radiation frequency.}
 \label{Directivity_Sigma}
\end{figure}

For the magnetic dipole emitter $\mathbf{m}_1$ the expressions (\ref{eq1}) and (\ref{eq2}) can be written as $P_{\rm ex}=\omega/2\mbox{Im}\left[\mathbf{m}_1^*\cdot \mathbf{H}_2(0)\right]$ and $\mathbf{m}_1\mathbf{H}_{2}(0)=\mathbf{m}_2\mathbf{H}_{1}(\mathbf{R})$, respectively. Performing all the same conversions, we arrive at a similar expression (\ref{eq4}) for the PF of a magnetic dipole emitter $F_{\rm rad}=3/2(\Sigma^2_m/D)$. This expression is similar to Eq. (\ref{eq4}) up to replacement of the LFEF $\Sigma_e$ by a local-field enhancement factor of \textit{a magnetic field} $\Sigma_m$.

\begin{figure*}
\includegraphics[width=1.0\textwidth]{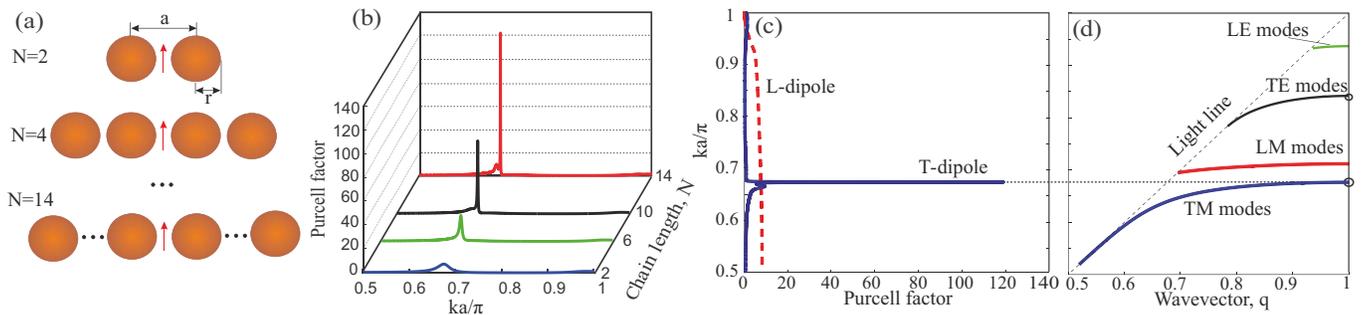}
\caption{(a) All-dielectric di-mer (N$=2$), exi-mer (N$=6$), ..., and dekatessera-mer (N$=14$) nanoantennas in form of chains with the same period $a=200$ nm. The dielectric constant of the nanoparticles is $\varepsilon=16$, the radii of nanoparticles are $r=70$ nm. (b) The dependences of Purcell factor for different chain length $N=2-14$ for T--dipole orientation on radiation frequency. (c) Spectrum of Purcell enhancement computed for $T$ and $L$ polarizations of the dipole emitter for finite chain consisting of $N=14$ nanoparticles. (d) Dispersion curves of eigenmodes in infinite dielectric chain.}\label{Fig2_theor}
\end{figure*}

\section{Large Purcell effect in all-dielectric nanostructures}

Our challenge is to apply the expression (\ref{eq4}) to achieve the high PF in dielectric nanostructures with a strong magnetic response. We want to utilize namely magnetic resonance in such structures. From the above analysis it is clear that such all-dielectric nanostructures should be symmetrical in respect to the emitter location. In such a structure it is necessary to excite preferably a quadrupole eigenmode that does not radiate in its orthogonal plane. Let us consider the simplest system of an all-dielectric dimer consisting of two dielectric nanoparticles with a high refractive index (such as a crystalline silicon Si). For analysis we choose a dielectric material with the dielectric constant of $\varepsilon=16$. The nanoparticle of this material with radius of $r=70$ nm has the magnetic dipole resonance at the frequency 463 THz. The distance between the centers of the nanoparticles is equal to $a=200$ nm. We put an electric dipole emitter (e.g. molecule, quantum dot, NV--center in nanodiamond) exactly in the middle between the nanoparticles orthogonally their axial axis (as presented in Fig.~\ref{Directivity_Sigma}(a), insert in the upper right corner). In this case, at the magnetic resonance frequency of the nanoparticles, their magnetic moments oscillate with a phase difference $\pi$, i.e. in antiphase. The radiation pattern of this system is a result of electric dipole (source) and magnetic quadrupole (all-dielectric dimer nanoantenna) radiation. The radiation directivity in ($\theta=\pi/2$, $\varphi=\pi/2$) direction (red solid curve) is strongly suppressed. The local-field enhancement factor (black dashed curve) under irradiation from the same direction, for this all-dielectric dimer nanoantenna \textit{even lower than 1}. Therefore, one might expect that the PF is also very low. However, its value reaches 10. In the Fig.~\ref{Directivity_Sigma}(b) the Purcell factor of all-dielectric dimer nanoantenna calculated by the formula (\ref{eq4}) (red curve) is presented. For comparison, the PF of this dimer nanoantenna obtained by the discrete dipole aproximation (DDA, blue crosses), and calculated through the method of input impedances~\cite{Krasnok_Purcell_2015} are presented. The frequency at which the maximum value of the PF of this dimer nanoantenna is observed is close to the frequency of the magnetic resonance of a single nanoparticle.

By increasing the number of particles $N$ by two [see Fig.~\ref{Fig2_theor}(a)], we observe a significant increase of PF from 10 ($N=2$) up to 120 ($N=14$). The PF of this nanoantennas have been calculated using the Green's function approach. The results are presented in the Fig.~\ref{Fig2_theor}(b), where the dimensionless frequency $\omega=ka$ is used. For example, in the Fig.~\ref{Fig2_theor}(c) the PF dependences on the frequency for the case $N=14$ for parallel (L--dipole, dashed curve) and perpendicular (T--dipole, solid curve) orientations of the dipole emitter are presented. The maximum value of the PF increases at the frequency 457 THz where each nanoparticle is polarized in antiphase in relation to its neighboring nanoparticles. The radiation of the nanostructure in such a coherent state can be classified as a classical analog of \textit{subradiance effect}~\cite{Pasquiou2014,Hommel_2007}.

\begin{figure*}[!t]
\includegraphics[width=1.0\textwidth]{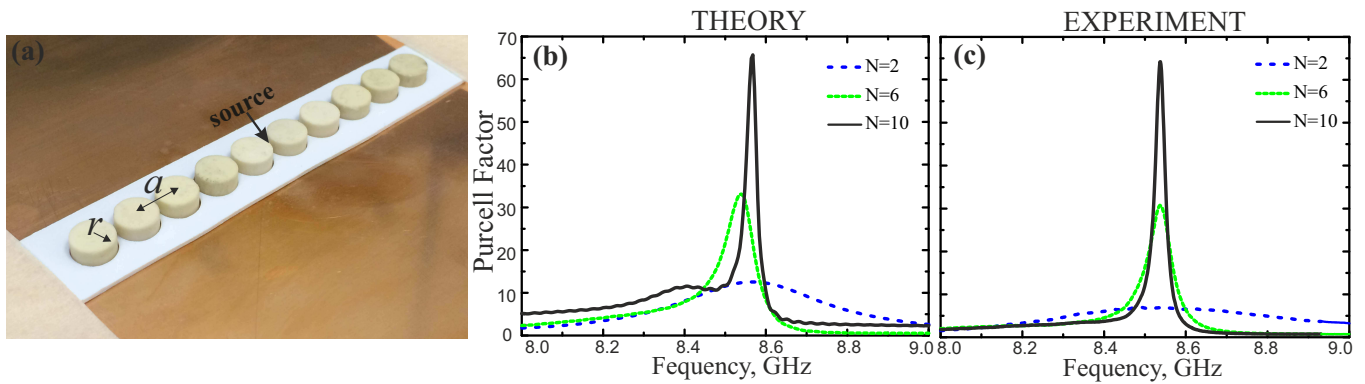}
\caption{Experimental verification of the Purcell enhancement. (a) The geometry of the experimental setup. (b) Numerical simulations of the Purcell factor for a chain of $N=2-10$ dielectric disks. (c) Experimentally measured Purcell factor by 3-mm dipole antenna placed in the center of the chain. The inset shows the geometry. Parameters of numerical simulations and experiment are: $r=4$ mm, $H=4$~mm, $a=5$~mm, and $\varepsilon=16$. The method of the Purcell factor measurement and calculation is taken from~\cite{Krasnok_Purcell_2015}.} \label{Fig3_exp}
\end{figure*}

First, we discuss an infinite chain. The dispersion properties of an infinite chain with period $a=200$ nm consisting of dielectric nanospheres with permittivity $\varepsilon=16$ and radius $r=70$ nm embedded in free space are illustrated by Fig.~\ref{Fig2_theor}(d). Here the dipole-dipole interaction model~\cite{Weber2004} have been used: each sphere is replaced by its electric and magnetic dipoles. The set of eigenmodes consists of four branches: transverse electric (TE), transverse magnetic (TM), longitudinal electric (LE), and longitudinal magnetic (LM). The corresponding dispersion curves $\om(q)$ are shown in Fig.~\ref{Fig2_theor}(d), where the dimensionless variables are used: normalized frequency $\om=ka/\pi$, where $k$ is the wave vector, and normalized quasi vector $q=\beta a/\pi$, where $\beta$ is the propagation constant. The leaky branches lying above the ''light line'' are not shown. The PF spectra for a dipole polarized along (L--dipole) and perpendicular (T--dipole) to the chain axis are shown in Fig.~\ref{Fig2_theor}(c) fore a case $N=14$. One can see that drastic PF enhancement is predicted for T--dipole at the frequency corresponding to TM--mode band edge. Having almost zero group velocity at the band edge the chain modes has divergent density of states due to \textit{van Hove singularity} in the infinite 1D structure, that may result in ultrahigh Purcell factor~\cite{Fink_2004}. However, high density of states is observed for TM--mode only, and is not observed for L--polarized dipole in the whole spectrum. The decision of this effect lies in the symmetry of the band edge modes. The phase shift between the dipole moments of the neighboring nanoparticles equals to $\pi$ at the band edge that makes these modes to be sufficiently ''dark''. The coupling between the dipole emitter and TE--, LE-- and LM--modes is suppressed due to the  \textit{symmetry mismatch}. It can be illustrated on the example of T--dipole and TE--mode: the electric fields induced by the oppositely polarized nanoparticles cancel each other in the point of the dipole emitter. Thus, the excitation of TE--mode by T-dipole is principally prohibited that also applies to LE-- and LM--modes. However, the TM--mode possesses the required symmetry and the electric fields generated by neighboring nanoparticles compliment each other, thus, the TM--mode is effectively excited. Having high density of states these modes give rise to the Purcell factor. Figure~\ref{Fig2_theor}(b) shows that this effect relates to collective excitation of dark mode. One can see that Purcell factor rapidly increases with the increase of chain length that is explained by forming mode structure similar to infinite chain when the conception of group velocity starts to be applicable.

\section{Experimental verification}

Since the fabrication and measurement of nanospheres operating in the optical frequency range is difficult, below we demonstrate the proof-of-concept experiment for the microwave frequency range, similar to earlier studies~\cite{Savelev2014_1}. We scale up all the dimensions and perform numerical simulations and experimental studies. We use MgO-TiO$_2$ ceramic disks with permittivity $\varepsilon=16$ and dielectric loss factor of $1{\rm e}^{-3}$. In order to confirm experimentally the frequency dependence of the Purcell factor, we measure directly the input impedance of an electrically short wire antenna exciting a chain of ceramic cylinders at microwaves [see Fig.~\ref{Fig3_exp}(a)]. We have considered a ceramic chain with the following parameters: the radius of the ceramic cylinder $r=4$ mm, the cylinder height $H=4$~mm, and the period of the chain $a=5$~mm. In such a measurement the Purcell factor is found as a ratio between the radiation resistance $Z$ of the antenna in presence of the structure and the radiation resistance $Z_0$ of the same antenna situated in free space~\cite{Krasnok_Purcell_2015}. The radiation resistance of small dipole antennas characterizes the radiated power, and it is equal to the real part of the radiation resistance measured directly in the feeding point. The ceramic cylinders used in the experiment exhibit their individual magnetic dipole resonances at 8.5 GHz. Therefore, to measure correctly the real part of the input impedance of an electrically short antenna (of the length $L \ll \lambda$) in the broad frequency range, we use a monopole over a metal ground plane [a mirror in Fig.~\ref{Fig3_exp}(a)]. Such a setup requires no balanced-to-unbalanced transformer, and it can be measured in a standard way with a single calibrated 50-Ohm coaxial port of vector network analyzer (Rohde\&Schwarz ZVB20). However, the heights of the antenna and the cylinders effectively double due to the metal ground plane. The ceramic cylinders are arranged in a chain with a thin perforated polyethylene holder, which does not affect the electromagnetic properties of the structure. The monopole antenna is formed by a 0.5 mm--thick core of the coaxial cable going through a hole in a copper sheet. The monopole height above the ground plane is chosen to be as small as 3 mm to avoid its own resonances in the range from 7 to 11 GHz.

The results of numerical simulations are presented in Fig.~\ref{Fig3_exp}(b). We observe a strong increase of the Purcell factor at frequency of 8.6 GHz. In Fig.~\ref{Fig3_exp}(c), we present the results of experimental measurements of the Purcell factor. We observe an excellent agreement between the experimental and numerical results. We notice that for a different orientation of the cylinders in the chain (the cylinders axis is oriented along the axis of the chain) we obtain much smaller values of the Purcell factor. This can be explained as follows. Although, the local electric field enhancement $\Sigma_e$ in the considered chain is less the radiation pattern has more suppressed side lobes. Therefore, the geometry of chain shown in Fig.~\ref{Fig3_exp}(a) provides high values of the Purcell factor, in accordance with Eq.(\ref{eq4}).

\section{Conclusions}

 We have proposed a novel approach for tailoring the Purcell factor by combining the local-field enhancement with radiation directivity. We have predicted that large values of the Purcell factor can be achieved even in radiating systems without large local-field enhancement but having a specific structure of the radiation patterns. Based on this approach, we have shown that in all-dielectric nanostructures the Purcell factor can be increased by two orders of magnitude in comparison with previously reported data. We have confirmed our theoretical predictions by a proof-of-principle microwave experiments with arrays of high-index subwavelength dielectric particles, suggesting similar effect for nanoscale structures due to scalability of the all-dielectric photonics.

\section*{Acknowledgements}

We thank E.~A. Nenasheva and D. Filonov for a technical help. This work was supported by the Ministry of Education and Science of the Russian Federation (project 14.584.21.0009 10), the Russian Foundation for Basic Research, and the Australian Research Council.


\begin{thebibliography}{35}%
\makeatletter
\providecommand \@ifxundefined [1]{%
 \@ifx{#1\undefined}
}%
\providecommand \@ifnum [1]{%
 \ifnum #1\expandafter \@firstoftwo
 \else \expandafter \@secondoftwo
 \fi
}%
\providecommand \@ifx [1]{%
 \ifx #1\expandafter \@firstoftwo
 \else \expandafter \@secondoftwo
 \fi
}%
\providecommand \natexlab [1]{#1}%
\providecommand \enquote  [1]{``#1''}%
\providecommand \bibnamefont  [1]{#1}%
\providecommand \bibfnamefont [1]{#1}%
\providecommand \citenamefont [1]{#1}%
\providecommand \href@noop [0]{\@secondoftwo}%
\providecommand \href [0]{\begingroup \@sanitize@url \@href}%
\providecommand \@href[1]{\@@startlink{#1}\@@href}%
\providecommand \@@href[1]{\endgroup#1\@@endlink}%
\providecommand \@sanitize@url [0]{\catcode `\\12\catcode `\$12\catcode
  `\&12\catcode `\#12\catcode `\^12\catcode `\_12\catcode `\%12\relax}%
\providecommand \@@startlink[1]{}%
\providecommand \@@endlink[0]{}%
\providecommand \url  [0]{\begingroup\@sanitize@url \@url }%
\providecommand \@url [1]{\endgroup\@href {#1}{\urlprefix }}%
\providecommand \urlprefix  [0]{URL }%
\providecommand \Eprint [0]{\href }%
\providecommand \doibase [0]{http://dx.doi.org/}%
\providecommand \selectlanguage [0]{\@gobble}%
\providecommand \bibinfo  [0]{\@secondoftwo}%
\providecommand \bibfield  [0]{\@secondoftwo}%
\providecommand \translation [1]{[#1]}%
\providecommand \BibitemOpen [0]{}%
\providecommand \bibitemStop [0]{}%
\providecommand \bibitemNoStop [0]{.\EOS\space}%
\providecommand \EOS [0]{\spacefactor3000\relax}%
\providecommand \BibitemShut  [1]{\csname bibitem#1\endcsname}%
\let\auto@bib@innerbib\@empty
\bibitem [{\citenamefont {Purcell}(1946)}]{Purcell_46}%
  \BibitemOpen
  \bibfield  {author} {\bibinfo {author} {\bibfnamefont {E.~M.}\ \bibnamefont
  {Purcell}},\ }\href@noop {} {\bibfield  {journal} {\bibinfo  {journal} {Phys.
  Rev.}\ }\textbf {\bibinfo {volume} {69}},\ \bibinfo {pages} {681} (\bibinfo
  {year} {1946})}\BibitemShut {NoStop}%
\bibitem [{\citenamefont {Sauvan}\ \emph {et~al.}(2013)\citenamefont {Sauvan},
  \citenamefont {Hugonin}, \citenamefont {Maksymov},\ and\ \citenamefont
  {Lalanne}}]{Sauvan2013a}%
  \BibitemOpen
  \bibfield  {author} {\bibinfo {author} {\bibfnamefont {C.}~\bibnamefont
  {Sauvan}}, \bibinfo {author} {\bibfnamefont {J.~P.}\ \bibnamefont {Hugonin}},
  \bibinfo {author} {\bibfnamefont {I.~S.}\ \bibnamefont {Maksymov}}, \ and\
  \bibinfo {author} {\bibfnamefont {P.}~\bibnamefont {Lalanne}},\ }\href
  {\doibase 10.1103/PhysRevLett.110.237401} {\bibfield  {journal} {\bibinfo
  {journal} {Physical Review Letters}\ }\textbf {\bibinfo {volume} {110}},\
  \bibinfo {pages} {237401} (\bibinfo {year} {2013})}\BibitemShut {NoStop}%
\bibitem [{\citenamefont {Nezhad}\ \emph {et~al.}(2010)\citenamefont {Nezhad},
  \citenamefont {Simic}, \citenamefont {Bondarenko}, \citenamefont {Slutsky},
  \citenamefont {Mizrahi}, \citenamefont {Feng}, \citenamefont {Lomakin},\ and\
  \citenamefont {Fainman}}]{Fainman_2010}%
  \BibitemOpen
  \bibfield  {author} {\bibinfo {author} {\bibfnamefont {M.~P.}\ \bibnamefont
  {Nezhad}}, \bibinfo {author} {\bibfnamefont {A.}~\bibnamefont {Simic}},
  \bibinfo {author} {\bibfnamefont {O.}~\bibnamefont {Bondarenko}}, \bibinfo
  {author} {\bibfnamefont {B.}~\bibnamefont {Slutsky}}, \bibinfo {author}
  {\bibfnamefont {A.}~\bibnamefont {Mizrahi}}, \bibinfo {author} {\bibfnamefont
  {L.}~\bibnamefont {Feng}}, \bibinfo {author} {\bibfnamefont {V.}~\bibnamefont
  {Lomakin}}, \ and\ \bibinfo {author} {\bibfnamefont {Y.}~\bibnamefont
  {Fainman}},\ }\href@noop {} {\bibfield  {journal} {\bibinfo  {journal}
  {Nature Photonics}\ }\textbf {\bibinfo {volume} {4}},\ \bibinfo {pages} {395}
  (\bibinfo {year} {2010})}\BibitemShut {NoStop}%
\bibitem [{\citenamefont {Frimmer}\ \emph {et~al.}(2011)\citenamefont
  {Frimmer}, \citenamefont {Chen},\ and\ \citenamefont
  {Koenderink}}]{Koenderink_PRL_11}%
  \BibitemOpen
  \bibfield  {author} {\bibinfo {author} {\bibfnamefont {M.}~\bibnamefont
  {Frimmer}}, \bibinfo {author} {\bibfnamefont {Y.}~\bibnamefont {Chen}}, \
  and\ \bibinfo {author} {\bibfnamefont {A.~F.}\ \bibnamefont {Koenderink}},\
  }\href@noop {} {\bibfield  {journal} {\bibinfo  {journal} {Physical Review
  Letters}\ }\textbf {\bibinfo {volume} {107}},\ \bibinfo {pages} {123602}
  (\bibinfo {year} {2011})}\BibitemShut {NoStop}%
\bibitem [{\citenamefont {Cosa}(2013)}]{Cosa_2013}%
  \BibitemOpen
  \bibfield  {author} {\bibinfo {author} {\bibfnamefont {G.}~\bibnamefont
  {Cosa}},\ }\href@noop {} {\bibfield  {journal} {\bibinfo  {journal} {Nature
  Chemistry}\ }\textbf {\bibinfo {volume} {5}},\ \bibinfo {pages} {159}
  (\bibinfo {year} {2013})}\BibitemShut {NoStop}%
\bibitem [{\citenamefont {Bermel}\ \emph {et~al.}(2007)\citenamefont {Bermel},
  \citenamefont {Rodriguez}, \citenamefont {Joannopoulos},\ and\ \citenamefont
  {Soljacic}}]{Soljacic_2007}%
  \BibitemOpen
  \bibfield  {author} {\bibinfo {author} {\bibfnamefont {P.}~\bibnamefont
  {Bermel}}, \bibinfo {author} {\bibfnamefont {A.}~\bibnamefont {Rodriguez}},
  \bibinfo {author} {\bibfnamefont {J.~D.}\ \bibnamefont {Joannopoulos}}, \
  and\ \bibinfo {author} {\bibfnamefont {M.}~\bibnamefont {Soljacic}},\
  }\href@noop {} {\bibfield  {journal} {\bibinfo  {journal} {Physical Review
  Letters}\ }\textbf {\bibinfo {volume} {99}},\ \bibinfo {pages} {053601}
  (\bibinfo {year} {2007})}\BibitemShut {NoStop}%
\bibitem [{\citenamefont {Todorov}\ \emph {et~al.}(2007)\citenamefont
  {Todorov}, \citenamefont {Sagnes}, \citenamefont {Abram},\ and\ \citenamefont
  {Minot}}]{Minot_2007}%
  \BibitemOpen
  \bibfield  {author} {\bibinfo {author} {\bibfnamefont {Y.}~\bibnamefont
  {Todorov}}, \bibinfo {author} {\bibfnamefont {I.}~\bibnamefont {Sagnes}},
  \bibinfo {author} {\bibfnamefont {I.}~\bibnamefont {Abram}}, \ and\ \bibinfo
  {author} {\bibfnamefont {C.}~\bibnamefont {Minot}},\ }\href@noop {}
  {\bibfield  {journal} {\bibinfo  {journal} {Physical Review Letters}\
  }\textbf {\bibinfo {volume} {99}},\ \bibinfo {pages} {223603} (\bibinfo
  {year} {2007})}\BibitemShut {NoStop}%
\bibitem [{\citenamefont {Beams}\ \emph {et~al.}(2013)\citenamefont {Beams},
  \citenamefont {Smith}, \citenamefont {Johnson}, \citenamefont {Oh},
  \citenamefont {Novotny},\ and\ \citenamefont
  {Vamivakas}}]{Vamivakas_NanoLetters_13}%
  \BibitemOpen
  \bibfield  {author} {\bibinfo {author} {\bibfnamefont {R.}~\bibnamefont
  {Beams}}, \bibinfo {author} {\bibfnamefont {D.}~\bibnamefont {Smith}},
  \bibinfo {author} {\bibfnamefont {T.~W.}\ \bibnamefont {Johnson}}, \bibinfo
  {author} {\bibfnamefont {S.-H.}\ \bibnamefont {Oh}}, \bibinfo {author}
  {\bibfnamefont {L.}~\bibnamefont {Novotny}}, \ and\ \bibinfo {author}
  {\bibfnamefont {A.~N.}\ \bibnamefont {Vamivakas}},\ }\href@noop {} {\bibfield
   {journal} {\bibinfo  {journal} {Nano Letters}\ }\textbf {\bibinfo {volume}
  {13}},\ \bibinfo {pages} {3807} (\bibinfo {year} {2013})}\BibitemShut
  {NoStop}%
\bibitem [{\citenamefont {Aigouy}\ \emph {et~al.}(2014)\citenamefont {Aigouy},
  \citenamefont {Caze}, \citenamefont {Gredin}, \citenamefont {Mortier},\ and\
  \citenamefont {Carminati}}]{Carminati_PRL_14}%
  \BibitemOpen
  \bibfield  {author} {\bibinfo {author} {\bibfnamefont {L.}~\bibnamefont
  {Aigouy}}, \bibinfo {author} {\bibfnamefont {A.}~\bibnamefont {Caze}},
  \bibinfo {author} {\bibfnamefont {P.}~\bibnamefont {Gredin}}, \bibinfo
  {author} {\bibfnamefont {M.}~\bibnamefont {Mortier}}, \ and\ \bibinfo
  {author} {\bibfnamefont {R.}~\bibnamefont {Carminati}},\ }\href@noop {}
  {\bibfield  {journal} {\bibinfo  {journal} {Physical Review Letters}\
  }\textbf {\bibinfo {volume} {113}},\ \bibinfo {pages} {076101} (\bibinfo
  {year} {2014})}\BibitemShut {NoStop}%
\bibitem [{\citenamefont {Acuna}\ \emph {et~al.}(2012)\citenamefont {Acuna},
  \citenamefont {Moller}, \citenamefont {Holzmeister}, \citenamefont {Beater},
  \citenamefont {Lalkens},\ and\ \citenamefont
  {Tinnefeld}}]{Tinnefeld_Science_2012}%
  \BibitemOpen
  \bibfield  {author} {\bibinfo {author} {\bibfnamefont {G.~P.}\ \bibnamefont
  {Acuna}}, \bibinfo {author} {\bibfnamefont {F.~M.}\ \bibnamefont {Moller}},
  \bibinfo {author} {\bibfnamefont {P.}~\bibnamefont {Holzmeister}}, \bibinfo
  {author} {\bibfnamefont {S.}~\bibnamefont {Beater}}, \bibinfo {author}
  {\bibfnamefont {B.}~\bibnamefont {Lalkens}}, \ and\ \bibinfo {author}
  {\bibfnamefont {P.}~\bibnamefont {Tinnefeld}},\ }\href@noop {} {\bibfield
  {journal} {\bibinfo  {journal} {Science}\ }\textbf {\bibinfo {volume}
  {338}},\ \bibinfo {pages} {506} (\bibinfo {year} {2012})}\BibitemShut
  {NoStop}%
\bibitem [{\citenamefont {Koenderink}(2010)}]{Koenderink2010}%
  \BibitemOpen
  \bibfield  {author} {\bibinfo {author} {\bibfnamefont {A.~F.}\ \bibnamefont
  {Koenderink}},\ }\href {http://www.ncbi.nlm.nih.gov/pubmed/21165139}
  {\bibfield  {journal} {\bibinfo  {journal} {Optics letters}\ }\textbf
  {\bibinfo {volume} {35}},\ \bibinfo {pages} {4208} (\bibinfo {year}
  {2010})}\BibitemShut {NoStop}%
\bibitem [{\citenamefont {Novotny}\ and\ \citenamefont
  {Hecht}(2006)}]{Novotny_Hecht_book}%
  \BibitemOpen
  \bibfield  {author} {\bibinfo {author} {\bibfnamefont {L.}~\bibnamefont
  {Novotny}}\ and\ \bibinfo {author} {\bibfnamefont {B.}~\bibnamefont
  {Hecht}},\ }\href@noop {} {\emph {\bibinfo {title} {Principles of
  Nano-Optics}}}\ (\bibinfo  {publisher} {Cambridge University Press},\
  \bibinfo {year} {2006})\BibitemShut {NoStop}%
\bibitem [{\citenamefont {Barrow}\ \emph {et~al.}(2014)\citenamefont {Barrow},
  \citenamefont {Rossouw}, \citenamefont {Funston}, \citenamefont {Botton},\
  and\ \citenamefont {Mulvaney}}]{Barrow2014}%
  \BibitemOpen
  \bibfield  {author} {\bibinfo {author} {\bibfnamefont {S.~J.}\ \bibnamefont
  {Barrow}}, \bibinfo {author} {\bibfnamefont {D.}~\bibnamefont {Rossouw}},
  \bibinfo {author} {\bibfnamefont {A.~M.}\ \bibnamefont {Funston}}, \bibinfo
  {author} {\bibfnamefont {G.~A.}\ \bibnamefont {Botton}}, \ and\ \bibinfo
  {author} {\bibfnamefont {P.}~\bibnamefont {Mulvaney}},\ }\href@noop {}
  {\bibfield  {journal} {\bibinfo  {journal} {Nano Lett.}\ }\textbf {\bibinfo
  {volume} {14}},\ \bibinfo {pages} {3799} (\bibinfo {year}
  {2014})}\BibitemShut {NoStop}%
\bibitem [{\citenamefont {Popa}\ and\ \citenamefont
  {Cummer}(2008)}]{Cummer_08}%
  \BibitemOpen
  \bibfield  {author} {\bibinfo {author} {\bibfnamefont {B.-I.}\ \bibnamefont
  {Popa}}\ and\ \bibinfo {author} {\bibfnamefont {S.~A.}\ \bibnamefont
  {Cummer}},\ }\href@noop {} {\bibfield  {journal} {\bibinfo  {journal}
  {Physical Review Letters}\ }\textbf {\bibinfo {volume} {100}},\ \bibinfo
  {pages} {207401} (\bibinfo {year} {2008})}\BibitemShut {NoStop}%
\bibitem [{\citenamefont {Ginn}\ and\ \citenamefont
  {Brener}(2012)}]{Brener_12}%
  \BibitemOpen
  \bibfield  {author} {\bibinfo {author} {\bibfnamefont {J.~C.}\ \bibnamefont
  {Ginn}}\ and\ \bibinfo {author} {\bibfnamefont {I.}~\bibnamefont {Brener}},\
  }\href@noop {} {\bibfield  {journal} {\bibinfo  {journal} {Physical Review
  Letters}\ }\textbf {\bibinfo {volume} {108}},\ \bibinfo {pages} {097402}
  (\bibinfo {year} {2012})}\BibitemShut {NoStop}%
\bibitem [{\citenamefont {Fu}\ \emph {et~al.}(2013)\citenamefont {Fu},
  \citenamefont {Kuznetsov}, \citenamefont {Miroshnichenko}, \citenamefont
  {Yu},\ and\ \citenamefont {Lukyanchuk}}]{Lukyanchuk13}%
  \BibitemOpen
  \bibfield  {author} {\bibinfo {author} {\bibfnamefont {Y.~H.}\ \bibnamefont
  {Fu}}, \bibinfo {author} {\bibfnamefont {A.~I.}\ \bibnamefont {Kuznetsov}},
  \bibinfo {author} {\bibfnamefont {A.~E.}\ \bibnamefont {Miroshnichenko}},
  \bibinfo {author} {\bibfnamefont {Y.~F.}\ \bibnamefont {Yu}}, \ and\ \bibinfo
  {author} {\bibfnamefont {B.}~\bibnamefont {Lukyanchuk}},\ }\href@noop {}
  {\bibfield  {journal} {\bibinfo  {journal} {Nature Communications}\ }\textbf
  {\bibinfo {volume} {4}},\ \bibinfo {pages} {1527} (\bibinfo {year}
  {2013})}\BibitemShut {NoStop}%
\bibitem [{\citenamefont {Miroshnichenko}\ and\ \citenamefont
  {Kivshar}(2012)}]{Miroshnichenko:NL:2012}%
  \BibitemOpen
  \bibfield  {author} {\bibinfo {author} {\bibfnamefont {A.~E.}\ \bibnamefont
  {Miroshnichenko}}\ and\ \bibinfo {author} {\bibfnamefont {Y.~S.}\
  \bibnamefont {Kivshar}},\ }\href {\doibase 10.1021/nl303927q} {\bibfield
  {journal} {\bibinfo  {journal} {Nano Letters}\ }\textbf {\bibinfo {volume}
  {12}},\ \bibinfo {pages} {6459} (\bibinfo {year} {2012})}\BibitemShut
  {NoStop}%
\bibitem [{\citenamefont {Savelev}\ \emph {et~al.}(2014)\citenamefont
  {Savelev}, \citenamefont {Filonov}, \citenamefont {Kapitanova}, \citenamefont
  {Krasnok}, \citenamefont {Miroshnichenko}, \citenamefont {Belov},\ and\
  \citenamefont {Kivshar}}]{Savelev2014_1}%
  \BibitemOpen
  \bibfield  {author} {\bibinfo {author} {\bibfnamefont {R.~S.}\ \bibnamefont
  {Savelev}}, \bibinfo {author} {\bibfnamefont {D.~S.}\ \bibnamefont
  {Filonov}}, \bibinfo {author} {\bibfnamefont {P.~V.}\ \bibnamefont
  {Kapitanova}}, \bibinfo {author} {\bibfnamefont {A.~E.}\ \bibnamefont
  {Krasnok}}, \bibinfo {author} {\bibfnamefont {A.~E.}\ \bibnamefont
  {Miroshnichenko}}, \bibinfo {author} {\bibfnamefont {P.~A.}\ \bibnamefont
  {Belov}}, \ and\ \bibinfo {author} {\bibfnamefont {Y.~S.}\ \bibnamefont
  {Kivshar}},\ }\href {\doibase 10.1063/1.4901264} {\bibfield  {journal}
  {\bibinfo  {journal} {Applied Physics Letters}\ }\textbf {\bibinfo {volume}
  {105}},\ \bibinfo {pages} {181116} (\bibinfo {year} {2014})}\BibitemShut
  {NoStop}%
\bibitem [{\citenamefont {Decker}\ \emph {et~al.}(2015)\citenamefont {Decker},
  \citenamefont {Staude}, \citenamefont {Falkner}, \citenamefont {Dominguez},
  \citenamefont {Neshev}, \citenamefont {Brener}, \citenamefont {Pertsch},\
  and\ \citenamefont {Kivshar}}]{Staude_15}%
  \BibitemOpen
  \bibfield  {author} {\bibinfo {author} {\bibfnamefont {M.}~\bibnamefont
  {Decker}}, \bibinfo {author} {\bibfnamefont {I.}~\bibnamefont {Staude}},
  \bibinfo {author} {\bibfnamefont {M.}~\bibnamefont {Falkner}}, \bibinfo
  {author} {\bibfnamefont {J.}~\bibnamefont {Dominguez}}, \bibinfo {author}
  {\bibfnamefont {D.~N.}\ \bibnamefont {Neshev}}, \bibinfo {author}
  {\bibfnamefont {I.}~\bibnamefont {Brener}}, \bibinfo {author} {\bibfnamefont
  {T.}~\bibnamefont {Pertsch}}, \ and\ \bibinfo {author} {\bibfnamefont
  {Y.~S.}\ \bibnamefont {Kivshar}},\ }\href@noop {} {\bibfield  {journal}
  {\bibinfo  {journal} {Advanced Optical Materials}\ }\textbf {\bibinfo
  {volume} {DOI: 10.1002/adom.201400584}} (\bibinfo {year} {2015})}\BibitemShut
  {NoStop}%
\bibitem [{\citenamefont {Bakker}\ \emph {et~al.}(2015)\citenamefont {Bakker},
  \citenamefont {Permyakov}, \citenamefont {Yu}, \citenamefont {Markovich},
  \citenamefont {Paniagua-Dominguez}, \citenamefont {Gonzaga}, \citenamefont
  {Samusev}, \citenamefont {Kivshar}, \citenamefont {Luk'yanchuk},\ and\
  \citenamefont {Kuznetsov}}]{Kuznetsov15}%
  \BibitemOpen
  \bibfield  {author} {\bibinfo {author} {\bibfnamefont {R.~M.}\ \bibnamefont
  {Bakker}}, \bibinfo {author} {\bibfnamefont {D.}~\bibnamefont {Permyakov}},
  \bibinfo {author} {\bibfnamefont {Y.~F.}\ \bibnamefont {Yu}}, \bibinfo
  {author} {\bibfnamefont {D.}~\bibnamefont {Markovich}}, \bibinfo {author}
  {\bibfnamefont {R.}~\bibnamefont {Paniagua-Dominguez}}, \bibinfo {author}
  {\bibfnamefont {L.}~\bibnamefont {Gonzaga}}, \bibinfo {author} {\bibfnamefont
  {A.}~\bibnamefont {Samusev}}, \bibinfo {author} {\bibfnamefont
  {Y.}~\bibnamefont {Kivshar}}, \bibinfo {author} {\bibfnamefont
  {B.}~\bibnamefont {Luk'yanchuk}}, \ and\ \bibinfo {author} {\bibfnamefont
  {A.~I.}\ \bibnamefont {Kuznetsov}},\ }\href@noop {} {\bibfield  {journal}
  {\bibinfo  {journal} {Nano Letters}\ }\textbf {\bibinfo {volume} {15}},\
  \bibinfo {pages} {2137} (\bibinfo {year} {2015})}\BibitemShut {NoStop}%
\bibitem [{\citenamefont {Noda}\ \emph {et~al.}(2007)\citenamefont {Noda},
  \citenamefont {Fujita},\ and\ \citenamefont {Asano}}]{Asano2007}%
  \BibitemOpen
  \bibfield  {author} {\bibinfo {author} {\bibfnamefont {S.}~\bibnamefont
  {Noda}}, \bibinfo {author} {\bibfnamefont {M.}~\bibnamefont {Fujita}}, \ and\
  \bibinfo {author} {\bibfnamefont {T.}~\bibnamefont {Asano}},\ }\href@noop {}
  {\bibfield  {journal} {\bibinfo  {journal} {Nature Photonics}\ }\textbf
  {\bibinfo {volume} {1}},\ \bibinfo {pages} {449 } (\bibinfo {year}
  {2007})}\BibitemShut {NoStop}%
\bibitem [{\citenamefont {Schmidt}\ \emph {et~al.}(2012)\citenamefont
  {Schmidt}, \citenamefont {Esteban}, \citenamefont {Saenz}, \citenamefont
  {Suarez-Lacalle}, \citenamefont {Mackowski},\ and\ \citenamefont
  {Aizpurua}}]{Schmidt_12}%
  \BibitemOpen
  \bibfield  {author} {\bibinfo {author} {\bibfnamefont {M.~K.}\ \bibnamefont
  {Schmidt}}, \bibinfo {author} {\bibfnamefont {R.}~\bibnamefont {Esteban}},
  \bibinfo {author} {\bibfnamefont {J.~J.}\ \bibnamefont {Saenz}}, \bibinfo
  {author} {\bibfnamefont {I.}~\bibnamefont {Suarez-Lacalle}}, \bibinfo
  {author} {\bibfnamefont {S.}~\bibnamefont {Mackowski}}, \ and\ \bibinfo
  {author} {\bibfnamefont {J.}~\bibnamefont {Aizpurua}},\ }\href@noop {}
  {\bibfield  {journal} {\bibinfo  {journal} {Optics Express}\ }\textbf
  {\bibinfo {volume} {20}},\ \bibinfo {pages} {13636} (\bibinfo {year}
  {2012})}\BibitemShut {NoStop}%
\bibitem [{\citenamefont {Krasnok}\ \emph {et~al.}(2012)\citenamefont
  {Krasnok}, \citenamefont {Miroshnichenko}, \citenamefont {Belov},\ and\
  \citenamefont {Kivshar}}]{KrasnokOE}%
  \BibitemOpen
  \bibfield  {author} {\bibinfo {author} {\bibfnamefont {A.~E.}\ \bibnamefont
  {Krasnok}}, \bibinfo {author} {\bibfnamefont {A.~E.}\ \bibnamefont
  {Miroshnichenko}}, \bibinfo {author} {\bibfnamefont {P.~A.}\ \bibnamefont
  {Belov}}, \ and\ \bibinfo {author} {\bibfnamefont {Y.~S.}\ \bibnamefont
  {Kivshar}},\ }\href@noop {} {\bibfield  {journal} {\bibinfo  {journal}
  {Optics Express}\ }\textbf {\bibinfo {volume} {20}},\ \bibinfo {pages}
  {20599} (\bibinfo {year} {2012})}\BibitemShut {NoStop}%
\bibitem [{\citenamefont {Schuller}\ and\ \citenamefont
  {Brongersma}(2009)}]{Schuller:OE:2009}%
  \BibitemOpen
  \bibfield  {author} {\bibinfo {author} {\bibfnamefont {J.~A.}\ \bibnamefont
  {Schuller}}\ and\ \bibinfo {author} {\bibfnamefont {M.~L.}\ \bibnamefont
  {Brongersma}},\ }\href {\doibase 10.1364/OE.17.024084} {\bibfield  {journal}
  {\bibinfo  {journal} {Opt. Express}\ }\textbf {\bibinfo {volume} {17}},\
  \bibinfo {pages} {24084} (\bibinfo {year} {2009})}\BibitemShut {NoStop}%
\bibitem [{\citenamefont {Wylie}\ and\ \citenamefont {Sipe}(1984)}]{Sipe_1984}%
  \BibitemOpen
  \bibfield  {author} {\bibinfo {author} {\bibfnamefont {J.~M.}\ \bibnamefont
  {Wylie}}\ and\ \bibinfo {author} {\bibfnamefont {J.~E.}\ \bibnamefont
  {Sipe}},\ }\href@noop {} {\bibfield  {journal} {\bibinfo  {journal} {Physical
  Review A}\ }\textbf {\bibinfo {volume} {30}},\ \bibinfo {pages} {1185}
  (\bibinfo {year} {1984})}\BibitemShut {NoStop}%
\bibitem [{\citenamefont {Bharadwaj}\ \emph {et~al.}(2009)\citenamefont
  {Bharadwaj}, \citenamefont {Deutsch},\ and\ \citenamefont
  {Novotny}}]{Novotny_09_AOP}%
  \BibitemOpen
  \bibfield  {author} {\bibinfo {author} {\bibfnamefont {P.}~\bibnamefont
  {Bharadwaj}}, \bibinfo {author} {\bibfnamefont {B.}~\bibnamefont {Deutsch}},
  \ and\ \bibinfo {author} {\bibfnamefont {L.}~\bibnamefont {Novotny}},\
  }\href@noop {} {\bibfield  {journal} {\bibinfo  {journal} {Advances in Optics
  and Photonics}\ }\textbf {\bibinfo {volume} {1}},\ \bibinfo {pages} {438}
  (\bibinfo {year} {2009})}\BibitemShut {NoStop}%
\bibitem [{\citenamefont {Balanis}(2005)}]{Balanis2005}%
  \BibitemOpen
  \bibfield  {author} {\bibinfo {author} {\bibfnamefont {C.~A.}\ \bibnamefont
  {Balanis}},\ }\href
  {http://www.amazon.co.uk/Antenna-Theory-Analysis-Constantine-Balanis/dp/047166782X}
  {\emph {\bibinfo {title} {Antenna Theory: Analysis and Design}}},\ \bibinfo
  {edition} {3rd}\ ed.\ (\bibinfo  {publisher} {Wiley-Blackwell},\ \bibinfo
  {year} {2005})\ p.\ \bibinfo {pages} {1136}\BibitemShut {NoStop}%
\bibitem [{\citenamefont {Nordlander}(2013)}]{magdark_5}%
  \BibitemOpen
  \bibfield  {author} {\bibinfo {author} {\bibfnamefont {P.}~\bibnamefont
  {Nordlander}},\ }\href@noop {} {\bibfield  {journal} {\bibinfo  {journal}
  {Nature Nanotechnology}\ }\textbf {\bibinfo {volume} {8}},\ \bibinfo {pages}
  {76} (\bibinfo {year} {2013})}\BibitemShut {NoStop}%
\bibitem [{\citenamefont {Yan}\ \emph {et~al.}(2015)\citenamefont {Yan},
  \citenamefont {Liu}, \citenamefont {Lin}, \citenamefont {Wang}, \citenamefont
  {Chen}, \citenamefont {Wang},\ and\ \citenamefont {Yang}}]{Yang2015}%
  \BibitemOpen
  \bibfield  {author} {\bibinfo {author} {\bibfnamefont {J.}~\bibnamefont
  {Yan}}, \bibinfo {author} {\bibfnamefont {P.}~\bibnamefont {Liu}}, \bibinfo
  {author} {\bibfnamefont {Z.}~\bibnamefont {Lin}}, \bibinfo {author}
  {\bibfnamefont {H.}~\bibnamefont {Wang}}, \bibinfo {author} {\bibfnamefont
  {H.}~\bibnamefont {Chen}}, \bibinfo {author} {\bibfnamefont {C.}~\bibnamefont
  {Wang}}, \ and\ \bibinfo {author} {\bibfnamefont {G.}~\bibnamefont {Yang}},\
  }\href@noop {} {\bibfield  {journal} {\bibinfo  {journal} {ACS Nano}\
  }\textbf {\bibinfo {volume} {9}},\ \bibinfo {pages} {2968} (\bibinfo {year}
  {2015})}\BibitemShut {NoStop}%
\bibitem [{\citenamefont {Krasnok}\ \emph {et~al.}(2014)\citenamefont
  {Krasnok}, \citenamefont {Simovski}, \citenamefont {Belov},\ and\
  \citenamefont {Kivshar}}]{KrasnokNanoscale}%
  \BibitemOpen
  \bibfield  {author} {\bibinfo {author} {\bibfnamefont {A.~E.}\ \bibnamefont
  {Krasnok}}, \bibinfo {author} {\bibfnamefont {C.~R.}\ \bibnamefont
  {Simovski}}, \bibinfo {author} {\bibfnamefont {P.~A.}\ \bibnamefont {Belov}},
  \ and\ \bibinfo {author} {\bibfnamefont {Y.~S.}\ \bibnamefont {Kivshar}},\
  }\href@noop {} {\bibfield  {journal} {\bibinfo  {journal} {Nanoscale}\
  }\textbf {\bibinfo {volume} {6}},\ \bibinfo {pages} {7354} (\bibinfo {year}
  {2014})}\BibitemShut {NoStop}%
\bibitem [{\citenamefont {Krasnok}\ \emph {et~al.}(2015)\citenamefont
  {Krasnok}, \citenamefont {Slobozhanyuk}, \citenamefont {Simovski},
  \citenamefont {Tretyakov}, \citenamefont {Poddubny}, \citenamefont
  {Miroshnichenko}, \citenamefont {Kivshar},\ and\ \citenamefont
  {Belov}}]{Krasnok_Purcell_2015}%
  \BibitemOpen
  \bibfield  {author} {\bibinfo {author} {\bibfnamefont {A.~E.}\ \bibnamefont
  {Krasnok}}, \bibinfo {author} {\bibfnamefont {A.~P.}\ \bibnamefont
  {Slobozhanyuk}}, \bibinfo {author} {\bibfnamefont {C.~R.}\ \bibnamefont
  {Simovski}}, \bibinfo {author} {\bibfnamefont {S.~A.}\ \bibnamefont
  {Tretyakov}}, \bibinfo {author} {\bibfnamefont {A.~N.}\ \bibnamefont
  {Poddubny}}, \bibinfo {author} {\bibfnamefont {A.~E.}\ \bibnamefont
  {Miroshnichenko}}, \bibinfo {author} {\bibfnamefont {Y.~S.}\ \bibnamefont
  {Kivshar}}, \ and\ \bibinfo {author} {\bibfnamefont {P.~A.}\ \bibnamefont
  {Belov}},\ }\href@noop {} {\bibfield  {journal} {\bibinfo  {journal} {arXiv
  preprint}\ }\textbf {\bibinfo {volume} {arXiv:1501.04834}} (\bibinfo {year}
  {2015})}\BibitemShut {NoStop}%
\bibitem [{\citenamefont {Pasquiou}(2014)}]{Pasquiou2014}%
  \BibitemOpen
  \bibfield  {author} {\bibinfo {author} {\bibfnamefont {B.}~\bibnamefont
  {Pasquiou}},\ }\href@noop {} {\bibfield  {journal} {\bibinfo  {journal}
  {Nature Physics}\ }\textbf {\bibinfo {volume} {11}},\ \bibinfo {pages} {14}
  (\bibinfo {year} {2014})}\BibitemShut {NoStop}%
\bibitem [{\citenamefont {Scheibner}\ \emph {et~al.}(2007)\citenamefont
  {Scheibner}, \citenamefont {Schmidt}, \citenamefont {Worschech},
  \citenamefont {Forchel}, \citenamefont {Bacher}, \citenamefont {Passow},\
  and\ \citenamefont {Hommel}}]{Hommel_2007}%
  \BibitemOpen
  \bibfield  {author} {\bibinfo {author} {\bibfnamefont {M.}~\bibnamefont
  {Scheibner}}, \bibinfo {author} {\bibfnamefont {T.}~\bibnamefont {Schmidt}},
  \bibinfo {author} {\bibfnamefont {L.}~\bibnamefont {Worschech}}, \bibinfo
  {author} {\bibfnamefont {A.}~\bibnamefont {Forchel}}, \bibinfo {author}
  {\bibfnamefont {G.}~\bibnamefont {Bacher}}, \bibinfo {author} {\bibfnamefont
  {T.}~\bibnamefont {Passow}}, \ and\ \bibinfo {author} {\bibfnamefont
  {D.}~\bibnamefont {Hommel}},\ }\href@noop {} {\bibfield  {journal} {\bibinfo
  {journal} {Nature Physics}\ }\textbf {\bibinfo {volume} {3}},\ \bibinfo
  {pages} {106} (\bibinfo {year} {2007})}\BibitemShut {NoStop}%
\bibitem [{\citenamefont {Weber}\ and\ \citenamefont {Ford}(2004)}]{Weber2004}%
  \BibitemOpen
  \bibfield  {author} {\bibinfo {author} {\bibfnamefont {W.~H.}\ \bibnamefont
  {Weber}}\ and\ \bibinfo {author} {\bibfnamefont {G.~W.}\ \bibnamefont
  {Ford}},\ }\href {\doibase 10.1103/PhysRevB.70.125429} {\bibfield  {journal}
  {\bibinfo  {journal} {Physical Review B}\ }\textbf {\bibinfo {volume} {70}},\
  \bibinfo {pages} {125429} (\bibinfo {year} {2004})}\BibitemShut {NoStop}%
\bibitem [{\citenamefont {Bermel}\ \emph {et~al.}(2004)\citenamefont {Bermel},
  \citenamefont {Joannopoulos},\ and\ \citenamefont {Fink}}]{Fink_2004}%
  \BibitemOpen
  \bibfield  {author} {\bibinfo {author} {\bibfnamefont {P.}~\bibnamefont
  {Bermel}}, \bibinfo {author} {\bibfnamefont {J.~D.}\ \bibnamefont
  {Joannopoulos}}, \ and\ \bibinfo {author} {\bibfnamefont {Y.}~\bibnamefont
  {Fink}},\ }\href@noop {} {\bibfield  {journal} {\bibinfo  {journal} {Physical
  Review B}\ }\textbf {\bibinfo {volume} {69}},\ \bibinfo {pages} {035316}
  (\bibinfo {year} {2004})}\BibitemShut {NoStop}%
\end{thebibliography}

%

\end{document}